\documentclass[aps,pra,twocolumn]{revtex4}

\usepackage{bm}

\begin{document}

\title{Vortex-enhanced alternating order around impurities in antiferromagnets}

\author{Ruggero Vaia}
\affiliation{Istituto dei Sistemi Complessi,
             Consiglio Nazionale delle Ricerche,
             via Madonna del Piano 10,
             I-50019 Sesto Fiorentino (FI), Italy}
\author{Alessandro Cuccoli}
\affiliation{Dipartimento di Fisica,
             Universit\`a di Firenze e Unit\`a CNISM,
             via G. Sansone 1,
             I-50019 Sesto Fiorentino (FI), Italy}

\begin{abstract}
It has been recently pointed out that the presence of a nonmagnetic
impurity in a Heisenberg antiferromagnet generates an alternating
order of the surrounding spins which is independent of temperature in
a wide range. Quantum Monte Carlo simulations in the two-dimensional
$S=1/2$ case confirmed this picture, but showed a counterintuitive
enhancement of the alternating order around the Kosterlitz-Thouless
transition. We propose here an explanation in terms of the effect of
vortex excitations.
\end{abstract}

\maketitle

In a study of the effects of impurities in Heisenberg
antiferromagnets, Eggert, Sylju{\aa}sen, Anfuso, and
Andres~\cite{EggertSAA2007} used an effective model in order to
derive a universal expression for the function
$\langle{n_z(r)}\rangle$ describing the local alternating
magnetization induced around a vacancy at site ${\bm{r}}\,{=}\,0$ by
a uniform magnetic field $\bm{B}$ applied along the $z$-axis.
Basically, $\langle{n_z(r)}\rangle$ behaves as $e^{-r/\lambda}$ and
the main point of interest lies in the fact that the characteristic
length ~$\lambda(T)$~ is actually {\em independent} of the
temperature $T$, as they also checked in a wide temperature range by
quantum Monte Carlo simulations of the two-dimensional
$S\,{=}\,\frac12$ Heisenberg antiferromagnet. In addition, they point
out an `{\it exotic effect}' observed in their simulations: when $T$
steps over the estimated Kosterlitz-Thouless temperature
$T_{\rm{KT}}$ the temperature behavior of the induced alternating
order is nonmonotonic, i.e., ~$\lambda(T)$~ first
\emph{increases} and displays a little bump before being
progressively reduced as non-linear thermal fluctuations rapidly
destroy long-range correlations. The last sentence of
Ref.~\cite{EggertSAA2007} claims that this feature is
`counterintuitive and calls for further investigation'. Here we aim
indeed at giving an explanation in terms of the effects of vortex
excitations.

In what follows all the symbols, if not explicitly defined, will have
the same meaning as in Ref.~\cite{EggertSAA2007}.

The ingredients of the effective model of Ref.~\cite{EggertSAA2007}
are the spin stiffness $\rho_{\rm{s}}$ and what we here call the
effective elastic constant
$G\,{=}\,{B^2}/{4zJ}\,{\equiv}\,G_{\rm{u}}$. The latter is obtained
by studying the stability of the
\emph{uniform} spin-flop configuration, and it comes out that
$\lambda\,{=}\,\sqrt{{\rho_{\rm{s}}}/G}$.

However, if we allow for a `vortex' configuration the in-plane
components are nonuniform, their orientation being described by a
local azimuthal angle $\varphi_{\bm{i}}$. By following the same path
of Ref.~\cite{EggertSAA2007}, after setting
\[
 (s_{\bm{i}}^x,s_{\bm{i}}^y)\,{=}\,\pm{s}\,\cos(\alpha{\pm}\delta)\,
 \big(\cos\varphi_{\bm{i}},\sin\varphi_{\bm{i}}\big)
\]
and $s_{\bm{i}}^z\,{=}\,s\sin(\alpha{\pm}\delta)$ for the two
sublattices ($\pm$), the single-vortex correction to the effective
energy
\[
 \delta{}E_{\rm{eff}}=
 {\cal{K}}\,(\cos2\alpha+\cos2\delta){\cal{L}}~,
\]
is obtained~\cite{note}; in the last equation we have set
${\cal{K}}\,{=}\,zJs^2/2$ and
\begin{eqnarray*}
 {\cal{L}}&=& \frac1{2zN}\sum_{\langle\bm{ij}\rangle}
 \big[1-\cos(\varphi_{\bm{i}}\,{-}\,\varphi_{{\bm{j}}})\big]
\\
 &=& \frac1{8N}\int_1^R d^2(r{-}r_0)\,|\nabla\varphi(r)|^2
 = \frac\pi{4N}\ln R~,
\end{eqnarray*}
where the second line follows from using the continuum limit for a
single vortex centered in ${\bm{r}}_0$, so that
$|\nabla\varphi(r)|=1/|{\bm{r}}\,{-}\,{\bm{r}}_0|$; $\pi{R^2}$
represents the lattice surface available for the free-vortex.
Therefore, the full effective energy becomes
\begin{eqnarray*}
 E_{\rm{v}} &=& -{\cal{K}}(1\,{-}\,2{\cal{L}})
 -2{\cal{K}}{\cal{L}}\sin^2\!\alpha
 + 2{\cal{K}}(1\,{-}\,{\cal{L}})\,\sin^2\!\delta
\\
 &&\hspace{30mm}-Bs\,\cos\alpha\,\sin\delta~;
\end{eqnarray*}
minimization with respect to the canting angle $\delta$ gives
$4{\cal{K}}(1\,{-}\,{\cal{L}})\,\sin\delta\,{=}\,Bs\,\cos\alpha$, and
the resulting effective energy acquires the same form obtained in
Ref.~\cite{EggertSAA2007} but for the effective elastic constant that
now reads:
\[
 G=\frac{G_{\rm{u}}}{1\,{-}\,{\cal{L}}}-2{\cal{K}}{\cal{L}}
 \equiv G_{\rm{v}}~,
\]
with a relative deviation from the uniform case given by
\[
 \frac{\delta{G}}G\simeq
 -\Big[\Big(\frac{2zJs}{B}\Big)^2\,{-}\,1\Big]{\cal{L}}~.
\]
For $T\,{<}\,T_{\rm{KT}}$ vortices are tightly bound in pairs and
$R\,{\simeq}\,1$, so that $G_{\rm{v}}\,{\simeq}\,G_{\rm{u}}$. At
$T_{\rm{KT}}$ the first vortex-antivortex pair unbinds and
$R\,{\simeq}\,\sqrt{N}$. A rapid deviation of $G$ from the uniform
value $G_{\rm{u}}$ appears and the effect reported in
Ref.~\cite{EggertSAA2007} can be reasonably observed. Indeed, the
parameter values used in the inset of Fig.~2 of
Ref.~\cite{EggertSAA2007}, $N\,{=}\,128^2$ and $B/J\,{=}\,0.2$, give
${\cal{L}}\,{\simeq}\,3.8{\cdot}10^{-4}$ and
${2zJs}/{B}\,{\simeq}\,20$, resulting in
$\delta{G}/G\,{\simeq}\,-0.15$ and eventually
$\delta\lambda/\lambda\,{\simeq}\,{+}\,0.075$, a variation that is
compatible with what was observed in~\cite{EggertSAA2007}. The
initial increase of the characteristic length $\lambda$ at
$T_{\rm{KT}}$ can thus be reasonably traced back to free-vortices
softening the elastic force that induces the spin-flop configuration.

A more quantitative description of this effect is apparently
complicated and beyond the scope of this note. Nevertheless, one can
figure out that the exponentially fast decrease of the in-plane
correlation length $\xi(T)$ above $T_{\rm{KT}}$ would very soon
impose an upper bound to $\lambda$ inducing its progressive decrease
as temperature is further raised.


\begin{thebibliography}{00}
\bibitem{note}
When there are more than one free vortex (or antivortex) in a dilute
gas approximation the respective contributions simply add, and
${\cal{L}}\to n {\cal{L}}$, $n$ being the total number of vortex-like
excitations.
\end{thebibliography}
\end{document}